\documentclass[10pt,conference]{IEEEtran}
\IEEEoverridecommandlockouts
\usepackage{cite}
\usepackage{amsmath,amssymb,amsfonts}
\usepackage{algorithmic}
\usepackage{graphicx}
\usepackage{textcomp}
\usepackage{xcolor}
\def\BibTeX{{\rm B\kern-.05em{\sc i\kern-.025em b}\kern-.08em
    T\kern-.1667em\lower.7ex\hbox{E}\kern-.125emX}}
    
\usepackage{xargs}                      
\usepackage[colorinlistoftodos,prependcaption,textsize=tiny]{todonotes}

\definecolor{bg}{rgb}{0.95,0.95,0.95}

\usepackage{bbm}

\usepackage{comment}
\usepackage{graphicx}
\usepackage{amsmath}
\usepackage{amssymb}
\usepackage{pifont}

\usepackage[american]{babel}
\usepackage{microtype}
\usepackage{subcaption}
\usepackage[ruled,,linesnumbered ]{algorithm2e}
\SetAlFnt{\small}
\SetAlCapFnt{\small}

\usepackage{booktabs}%
\usepackage{color}%
\usepackage{xcolor}%
\definecolor{graybg}{rgb}{0.95,0.95,0.95}
\usepackage[draft]{minted}

\usepackage[flushleft]{threeparttable} 
\usepackage[normalem]{ulem}
\usepackage{tabularx}

\newcount\DraftStatus  
\usepackage{color}
\definecolor{darkgreen}{rgb}{0,0.5,0}
\definecolor{purple}{rgb}{1,0,1}
\definecolor{todocolor}{rgb}{0.9,0.1,0.1}
\definecolor{hycolor}{rgb}{0.7,0.7,0.3}
\definecolor{fixcolor}{rgb}{0.1,0.7,0.3}

\newcommand{\draftnote}[2]{\ifnum\DraftStatus=1
	\marginpar{
		\tiny\raggedright
		\hbadness=10000
		\def\baselinestretch{0.8}
		\textcolor{#1}{\textsf{\hspace{0pt}#2}}}
	\fi}

\newcommand{\tool}{\textsc{NaturalCC}\xspace}

\makeatletter
\def\PYG@reset{\let\PYG@it=\relax \let\PYG@bf=\relax%
	\let\PYG@ul=\relax \let\PYG@tc=\relax%
	\let\PYG@bc=\relax \let\PYG@ff=\relax}
\def\PYG@tok#1{\csname PYG@tok@#1\endcsname}
\def\PYG@toks#1+{\ifx\relax#1\empty\else%
	\PYG@tok{#1}\expandafter\PYG@toks\fi}
\def\PYG@do#1{\PYG@bc{\PYG@tc{\PYG@ul{%
				\PYG@it{\PYG@bf{\PYG@ff{#1}}}}}}}
\def\PYG#1#2{\PYG@reset\PYG@toks#1+\relax+\PYG@do{#2}}

\expandafter\def\csname PYG@tok@w\endcsname{\def\PYG@tc##1{\textcolor[rgb]{0.73,0.73,0.73}{##1}}}
\expandafter\def\csname PYG@tok@c\endcsname{\let\PYG@it=\textit\def\PYG@tc##1{\textcolor[rgb]{0.25,0.50,0.50}{##1}}}
\expandafter\def\csname PYG@tok@cp\endcsname{\def\PYG@tc##1{\textcolor[rgb]{0.74,0.48,0.00}{##1}}}
\expandafter\def\csname PYG@tok@k\endcsname{\let\PYG@bf=\textbf\def\PYG@tc##1{\textcolor[rgb]{0.00,0.50,0.00}{##1}}}
\expandafter\def\csname PYG@tok@kp\endcsname{\def\PYG@tc##1{\textcolor[rgb]{0.00,0.50,0.00}{##1}}}
\expandafter\def\csname PYG@tok@kt\endcsname{\def\PYG@tc##1{\textcolor[rgb]{0.69,0.00,0.25}{##1}}}
\expandafter\def\csname PYG@tok@o\endcsname{\def\PYG@tc##1{\textcolor[rgb]{0.40,0.40,0.40}{##1}}}
\expandafter\def\csname PYG@tok@ow\endcsname{\let\PYG@bf=\textbf\def\PYG@tc##1{\textcolor[rgb]{0.67,0.13,1.00}{##1}}}
\expandafter\def\csname PYG@tok@nb\endcsname{\def\PYG@tc##1{\textcolor[rgb]{0.00,0.50,0.00}{##1}}}
\expandafter\def\csname PYG@tok@nf\endcsname{\def\PYG@tc##1{\textcolor[rgb]{0.00,0.00,1.00}{##1}}}
\expandafter\def\csname PYG@tok@nc\endcsname{\let\PYG@bf=\textbf\def\PYG@tc##1{\textcolor[rgb]{0.00,0.00,1.00}{##1}}}
\expandafter\def\csname PYG@tok@nn\endcsname{\let\PYG@bf=\textbf\def\PYG@tc##1{\textcolor[rgb]{0.00,0.00,1.00}{##1}}}
\expandafter\def\csname PYG@tok@ne\endcsname{\let\PYG@bf=\textbf\def\PYG@tc##1{\textcolor[rgb]{0.82,0.25,0.23}{##1}}}
\expandafter\def\csname PYG@tok@nv\endcsname{\def\PYG@tc##1{\textcolor[rgb]{0.10,0.09,0.49}{##1}}}
\expandafter\def\csname PYG@tok@no\endcsname{\def\PYG@tc##1{\textcolor[rgb]{0.53,0.00,0.00}{##1}}}
\expandafter\def\csname PYG@tok@nl\endcsname{\def\PYG@tc##1{\textcolor[rgb]{0.63,0.63,0.00}{##1}}}
\expandafter\def\csname PYG@tok@ni\endcsname{\let\PYG@bf=\textbf\def\PYG@tc##1{\textcolor[rgb]{0.60,0.60,0.60}{##1}}}
\expandafter\def\csname PYG@tok@na\endcsname{\def\PYG@tc##1{\textcolor[rgb]{0.49,0.56,0.16}{##1}}}
\expandafter\def\csname PYG@tok@nt\endcsname{\let\PYG@bf=\textbf\def\PYG@tc##1{\textcolor[rgb]{0.00,0.50,0.00}{##1}}}
\expandafter\def\csname PYG@tok@nd\endcsname{\def\PYG@tc##1{\textcolor[rgb]{0.67,0.13,1.00}{##1}}}
\expandafter\def\csname PYG@tok@s\endcsname{\def\PYG@tc##1{\textcolor[rgb]{0.73,0.13,0.13}{##1}}}
\expandafter\def\csname PYG@tok@sd\endcsname{\let\PYG@it=\textit\def\PYG@tc##1{\textcolor[rgb]{0.73,0.13,0.13}{##1}}}
\expandafter\def\csname PYG@tok@si\endcsname{\let\PYG@bf=\textbf\def\PYG@tc##1{\textcolor[rgb]{0.73,0.40,0.53}{##1}}}
\expandafter\def\csname PYG@tok@se\endcsname{\let\PYG@bf=\textbf\def\PYG@tc##1{\textcolor[rgb]{0.73,0.40,0.13}{##1}}}
\expandafter\def\csname PYG@tok@sr\endcsname{\def\PYG@tc##1{\textcolor[rgb]{0.73,0.40,0.53}{##1}}}
\expandafter\def\csname PYG@tok@ss\endcsname{\def\PYG@tc##1{\textcolor[rgb]{0.10,0.09,0.49}{##1}}}
\expandafter\def\csname PYG@tok@sx\endcsname{\def\PYG@tc##1{\textcolor[rgb]{0.00,0.50,0.00}{##1}}}
\expandafter\def\csname PYG@tok@m\endcsname{\def\PYG@tc##1{\textcolor[rgb]{0.40,0.40,0.40}{##1}}}
\expandafter\def\csname PYG@tok@gh\endcsname{\let\PYG@bf=\textbf\def\PYG@tc##1{\textcolor[rgb]{0.00,0.00,0.50}{##1}}}
\expandafter\def\csname PYG@tok@gu\endcsname{\let\PYG@bf=\textbf\def\PYG@tc##1{\textcolor[rgb]{0.50,0.00,0.50}{##1}}}
\expandafter\def\csname PYG@tok@gd\endcsname{\def\PYG@tc##1{\textcolor[rgb]{0.63,0.00,0.00}{##1}}}
\expandafter\def\csname PYG@tok@gi\endcsname{\def\PYG@tc##1{\textcolor[rgb]{0.00,0.63,0.00}{##1}}}
\expandafter\def\csname PYG@tok@gr\endcsname{\def\PYG@tc##1{\textcolor[rgb]{1.00,0.00,0.00}{##1}}}
\expandafter\def\csname PYG@tok@ge\endcsname{\let\PYG@it=\textit}
\expandafter\def\csname PYG@tok@gs\endcsname{\let\PYG@bf=\textbf}
\expandafter\def\csname PYG@tok@gp\endcsname{\let\PYG@bf=\textbf\def\PYG@tc##1{\textcolor[rgb]{0.00,0.00,0.50}{##1}}}
\expandafter\def\csname PYG@tok@go\endcsname{\def\PYG@tc##1{\textcolor[rgb]{0.53,0.53,0.53}{##1}}}
\expandafter\def\csname PYG@tok@gt\endcsname{\def\PYG@tc##1{\textcolor[rgb]{0.00,0.27,0.87}{##1}}}
\expandafter\def\csname PYG@tok@err\endcsname{\def\PYG@bc##1{\setlength{\fboxsep}{0pt}\fcolorbox[rgb]{1.00,0.00,0.00}{1,1,1}{\strut ##1}}}
\expandafter\def\csname PYG@tok@kc\endcsname{\let\PYG@bf=\textbf\def\PYG@tc##1{\textcolor[rgb]{0.00,0.50,0.00}{##1}}}
\expandafter\def\csname PYG@tok@kd\endcsname{\let\PYG@bf=\textbf\def\PYG@tc##1{\textcolor[rgb]{0.00,0.50,0.00}{##1}}}
\expandafter\def\csname PYG@tok@kn\endcsname{\let\PYG@bf=\textbf\def\PYG@tc##1{\textcolor[rgb]{0.00,0.50,0.00}{##1}}}
\expandafter\def\csname PYG@tok@kr\endcsname{\let\PYG@bf=\textbf\def\PYG@tc##1{\textcolor[rgb]{0.00,0.50,0.00}{##1}}}
\expandafter\def\csname PYG@tok@bp\endcsname{\def\PYG@tc##1{\textcolor[rgb]{0.00,0.50,0.00}{##1}}}
\expandafter\def\csname PYG@tok@fm\endcsname{\def\PYG@tc##1{\textcolor[rgb]{0.00,0.00,1.00}{##1}}}
\expandafter\def\csname PYG@tok@vc\endcsname{\def\PYG@tc##1{\textcolor[rgb]{0.10,0.09,0.49}{##1}}}
\expandafter\def\csname PYG@tok@vg\endcsname{\def\PYG@tc##1{\textcolor[rgb]{0.10,0.09,0.49}{##1}}}
\expandafter\def\csname PYG@tok@vi\endcsname{\def\PYG@tc##1{\textcolor[rgb]{0.10,0.09,0.49}{##1}}}
\expandafter\def\csname PYG@tok@vm\endcsname{\def\PYG@tc##1{\textcolor[rgb]{0.10,0.09,0.49}{##1}}}
\expandafter\def\csname PYG@tok@sa\endcsname{\def\PYG@tc##1{\textcolor[rgb]{0.73,0.13,0.13}{##1}}}
\expandafter\def\csname PYG@tok@sb\endcsname{\def\PYG@tc##1{\textcolor[rgb]{0.73,0.13,0.13}{##1}}}
\expandafter\def\csname PYG@tok@sc\endcsname{\def\PYG@tc##1{\textcolor[rgb]{0.73,0.13,0.13}{##1}}}
\expandafter\def\csname PYG@tok@dl\endcsname{\def\PYG@tc##1{\textcolor[rgb]{0.73,0.13,0.13}{##1}}}
\expandafter\def\csname PYG@tok@s2\endcsname{\def\PYG@tc##1{\textcolor[rgb]{0.73,0.13,0.13}{##1}}}
\expandafter\def\csname PYG@tok@sh\endcsname{\def\PYG@tc##1{\textcolor[rgb]{0.73,0.13,0.13}{##1}}}
\expandafter\def\csname PYG@tok@s1\endcsname{\def\PYG@tc##1{\textcolor[rgb]{0.73,0.13,0.13}{##1}}}
\expandafter\def\csname PYG@tok@mb\endcsname{\def\PYG@tc##1{\textcolor[rgb]{0.40,0.40,0.40}{##1}}}
\expandafter\def\csname PYG@tok@mf\endcsname{\def\PYG@tc##1{\textcolor[rgb]{0.40,0.40,0.40}{##1}}}
\expandafter\def\csname PYG@tok@mh\endcsname{\def\PYG@tc##1{\textcolor[rgb]{0.40,0.40,0.40}{##1}}}
\expandafter\def\csname PYG@tok@mi\endcsname{\def\PYG@tc##1{\textcolor[rgb]{0.40,0.40,0.40}{##1}}}
\expandafter\def\csname PYG@tok@il\endcsname{\def\PYG@tc##1{\textcolor[rgb]{0.40,0.40,0.40}{##1}}}
\expandafter\def\csname PYG@tok@mo\endcsname{\def\PYG@tc##1{\textcolor[rgb]{0.40,0.40,0.40}{##1}}}
\expandafter\def\csname PYG@tok@ch\endcsname{\let\PYG@it=\textit\def\PYG@tc##1{\textcolor[rgb]{0.25,0.50,0.50}{##1}}}
\expandafter\def\csname PYG@tok@cm\endcsname{\let\PYG@it=\textit\def\PYG@tc##1{\textcolor[rgb]{0.25,0.50,0.50}{##1}}}
\expandafter\def\csname PYG@tok@cpf\endcsname{\let\PYG@it=\textit\def\PYG@tc##1{\textcolor[rgb]{0.25,0.50,0.50}{##1}}}
\expandafter\def\csname PYG@tok@c1\endcsname{\let\PYG@it=\textit\def\PYG@tc##1{\textcolor[rgb]{0.25,0.50,0.50}{##1}}}
\expandafter\def\csname PYG@tok@cs\endcsname{\let\PYG@it=\textit\def\PYG@tc##1{\textcolor[rgb]{0.25,0.50,0.50}{##1}}}

\def\PYGZus{\char`\_}

\def\PYGZlt{\char`\<}
\def\PYGZgt{\char`\>}
\def\PYGZsh{\char`\#}

\def\PYGZdl{\char`\$}
\def\PYGZhy{\char`\-}
\def\PYGZsq{\char`\'}

\makeatother

\makeatletter
\def\PYGdefault@reset{\let\PYGdefault@it=\relax \let\PYGdefault@bf=\relax%
	\let\PYGdefault@ul=\relax \let\PYGdefault@tc=\relax%
	\let\PYGdefault@bc=\relax \let\PYGdefault@ff=\relax}
\def\PYGdefault@tok#1{\csname PYGdefault@tok@#1\endcsname}
\def\PYGdefault@toks#1+{\ifx\relax#1\empty\else%
	\PYGdefault@tok{#1}\expandafter\PYGdefault@toks\fi}
\def\PYGdefault@do#1{\PYGdefault@bc{\PYGdefault@tc{\PYGdefault@ul{%
				\PYGdefault@it{\PYGdefault@bf{\PYGdefault@ff{#1}}}}}}}
\def\PYGdefault#1#2{\PYGdefault@reset\PYGdefault@toks#1+\relax+\PYGdefault@do{#2}}

\expandafter\def\csname PYGdefault@tok@w\endcsname{\def\PYGdefault@tc##1{\textcolor[rgb]{0.73,0.73,0.73}{##1}}}
\expandafter\def\csname PYGdefault@tok@c\endcsname{\let\PYGdefault@it=\textit\def\PYGdefault@tc##1{\textcolor[rgb]{0.25,0.50,0.50}{##1}}}
\expandafter\def\csname PYGdefault@tok@cp\endcsname{\def\PYGdefault@tc##1{\textcolor[rgb]{0.74,0.48,0.00}{##1}}}
\expandafter\def\csname PYGdefault@tok@k\endcsname{\let\PYGdefault@bf=\textbf\def\PYGdefault@tc##1{\textcolor[rgb]{0.00,0.50,0.00}{##1}}}
\expandafter\def\csname PYGdefault@tok@kp\endcsname{\def\PYGdefault@tc##1{\textcolor[rgb]{0.00,0.50,0.00}{##1}}}
\expandafter\def\csname PYGdefault@tok@kt\endcsname{\def\PYGdefault@tc##1{\textcolor[rgb]{0.69,0.00,0.25}{##1}}}
\expandafter\def\csname PYGdefault@tok@o\endcsname{\def\PYGdefault@tc##1{\textcolor[rgb]{0.40,0.40,0.40}{##1}}}
\expandafter\def\csname PYGdefault@tok@ow\endcsname{\let\PYGdefault@bf=\textbf\def\PYGdefault@tc##1{\textcolor[rgb]{0.67,0.13,1.00}{##1}}}
\expandafter\def\csname PYGdefault@tok@nb\endcsname{\def\PYGdefault@tc##1{\textcolor[rgb]{0.00,0.50,0.00}{##1}}}
\expandafter\def\csname PYGdefault@tok@nf\endcsname{\def\PYGdefault@tc##1{\textcolor[rgb]{0.00,0.00,1.00}{##1}}}
\expandafter\def\csname PYGdefault@tok@nc\endcsname{\let\PYGdefault@bf=\textbf\def\PYGdefault@tc##1{\textcolor[rgb]{0.00,0.00,1.00}{##1}}}
\expandafter\def\csname PYGdefault@tok@nn\endcsname{\let\PYGdefault@bf=\textbf\def\PYGdefault@tc##1{\textcolor[rgb]{0.00,0.00,1.00}{##1}}}
\expandafter\def\csname PYGdefault@tok@ne\endcsname{\let\PYGdefault@bf=\textbf\def\PYGdefault@tc##1{\textcolor[rgb]{0.82,0.25,0.23}{##1}}}
\expandafter\def\csname PYGdefault@tok@nv\endcsname{\def\PYGdefault@tc##1{\textcolor[rgb]{0.10,0.09,0.49}{##1}}}
\expandafter\def\csname PYGdefault@tok@no\endcsname{\def\PYGdefault@tc##1{\textcolor[rgb]{0.53,0.00,0.00}{##1}}}
\expandafter\def\csname PYGdefault@tok@nl\endcsname{\def\PYGdefault@tc##1{\textcolor[rgb]{0.63,0.63,0.00}{##1}}}
\expandafter\def\csname PYGdefault@tok@ni\endcsname{\let\PYGdefault@bf=\textbf\def\PYGdefault@tc##1{\textcolor[rgb]{0.60,0.60,0.60}{##1}}}
\expandafter\def\csname PYGdefault@tok@na\endcsname{\def\PYGdefault@tc##1{\textcolor[rgb]{0.49,0.56,0.16}{##1}}}
\expandafter\def\csname PYGdefault@tok@nt\endcsname{\let\PYGdefault@bf=\textbf\def\PYGdefault@tc##1{\textcolor[rgb]{0.00,0.50,0.00}{##1}}}
\expandafter\def\csname PYGdefault@tok@nd\endcsname{\def\PYGdefault@tc##1{\textcolor[rgb]{0.67,0.13,1.00}{##1}}}
\expandafter\def\csname PYGdefault@tok@s\endcsname{\def\PYGdefault@tc##1{\textcolor[rgb]{0.73,0.13,0.13}{##1}}}
\expandafter\def\csname PYGdefault@tok@sd\endcsname{\let\PYGdefault@it=\textit\def\PYGdefault@tc##1{\textcolor[rgb]{0.73,0.13,0.13}{##1}}}
\expandafter\def\csname PYGdefault@tok@si\endcsname{\let\PYGdefault@bf=\textbf\def\PYGdefault@tc##1{\textcolor[rgb]{0.73,0.40,0.53}{##1}}}
\expandafter\def\csname PYGdefault@tok@se\endcsname{\let\PYGdefault@bf=\textbf\def\PYGdefault@tc##1{\textcolor[rgb]{0.73,0.40,0.13}{##1}}}
\expandafter\def\csname PYGdefault@tok@sr\endcsname{\def\PYGdefault@tc##1{\textcolor[rgb]{0.73,0.40,0.53}{##1}}}
\expandafter\def\csname PYGdefault@tok@ss\endcsname{\def\PYGdefault@tc##1{\textcolor[rgb]{0.10,0.09,0.49}{##1}}}
\expandafter\def\csname PYGdefault@tok@sx\endcsname{\def\PYGdefault@tc##1{\textcolor[rgb]{0.00,0.50,0.00}{##1}}}
\expandafter\def\csname PYGdefault@tok@m\endcsname{\def\PYGdefault@tc##1{\textcolor[rgb]{0.40,0.40,0.40}{##1}}}
\expandafter\def\csname PYGdefault@tok@gh\endcsname{\let\PYGdefault@bf=\textbf\def\PYGdefault@tc##1{\textcolor[rgb]{0.00,0.00,0.50}{##1}}}
\expandafter\def\csname PYGdefault@tok@gu\endcsname{\let\PYGdefault@bf=\textbf\def\PYGdefault@tc##1{\textcolor[rgb]{0.50,0.00,0.50}{##1}}}
\expandafter\def\csname PYGdefault@tok@gd\endcsname{\def\PYGdefault@tc##1{\textcolor[rgb]{0.63,0.00,0.00}{##1}}}
\expandafter\def\csname PYGdefault@tok@gi\endcsname{\def\PYGdefault@tc##1{\textcolor[rgb]{0.00,0.63,0.00}{##1}}}
\expandafter\def\csname PYGdefault@tok@gr\endcsname{\def\PYGdefault@tc##1{\textcolor[rgb]{1.00,0.00,0.00}{##1}}}
\expandafter\def\csname PYGdefault@tok@ge\endcsname{\let\PYGdefault@it=\textit}
\expandafter\def\csname PYGdefault@tok@gs\endcsname{\let\PYGdefault@bf=\textbf}
\expandafter\def\csname PYGdefault@tok@gp\endcsname{\let\PYGdefault@bf=\textbf\def\PYGdefault@tc##1{\textcolor[rgb]{0.00,0.00,0.50}{##1}}}
\expandafter\def\csname PYGdefault@tok@go\endcsname{\def\PYGdefault@tc##1{\textcolor[rgb]{0.53,0.53,0.53}{##1}}}
\expandafter\def\csname PYGdefault@tok@gt\endcsname{\def\PYGdefault@tc##1{\textcolor[rgb]{0.00,0.27,0.87}{##1}}}
\expandafter\def\csname PYGdefault@tok@err\endcsname{\def\PYGdefault@bc##1{\setlength{\fboxsep}{0pt}\fcolorbox[rgb]{1.00,0.00,0.00}{1,1,1}{\strut ##1}}}
\expandafter\def\csname PYGdefault@tok@kc\endcsname{\let\PYGdefault@bf=\textbf\def\PYGdefault@tc##1{\textcolor[rgb]{0.00,0.50,0.00}{##1}}}
\expandafter\def\csname PYGdefault@tok@kd\endcsname{\let\PYGdefault@bf=\textbf\def\PYGdefault@tc##1{\textcolor[rgb]{0.00,0.50,0.00}{##1}}}
\expandafter\def\csname PYGdefault@tok@kn\endcsname{\let\PYGdefault@bf=\textbf\def\PYGdefault@tc##1{\textcolor[rgb]{0.00,0.50,0.00}{##1}}}
\expandafter\def\csname PYGdefault@tok@kr\endcsname{\let\PYGdefault@bf=\textbf\def\PYGdefault@tc##1{\textcolor[rgb]{0.00,0.50,0.00}{##1}}}
\expandafter\def\csname PYGdefault@tok@bp\endcsname{\def\PYGdefault@tc##1{\textcolor[rgb]{0.00,0.50,0.00}{##1}}}
\expandafter\def\csname PYGdefault@tok@fm\endcsname{\def\PYGdefault@tc##1{\textcolor[rgb]{0.00,0.00,1.00}{##1}}}
\expandafter\def\csname PYGdefault@tok@vc\endcsname{\def\PYGdefault@tc##1{\textcolor[rgb]{0.10,0.09,0.49}{##1}}}
\expandafter\def\csname PYGdefault@tok@vg\endcsname{\def\PYGdefault@tc##1{\textcolor[rgb]{0.10,0.09,0.49}{##1}}}
\expandafter\def\csname PYGdefault@tok@vi\endcsname{\def\PYGdefault@tc##1{\textcolor[rgb]{0.10,0.09,0.49}{##1}}}
\expandafter\def\csname PYGdefault@tok@vm\endcsname{\def\PYGdefault@tc##1{\textcolor[rgb]{0.10,0.09,0.49}{##1}}}
\expandafter\def\csname PYGdefault@tok@sa\endcsname{\def\PYGdefault@tc##1{\textcolor[rgb]{0.73,0.13,0.13}{##1}}}
\expandafter\def\csname PYGdefault@tok@sb\endcsname{\def\PYGdefault@tc##1{\textcolor[rgb]{0.73,0.13,0.13}{##1}}}
\expandafter\def\csname PYGdefault@tok@sc\endcsname{\def\PYGdefault@tc##1{\textcolor[rgb]{0.73,0.13,0.13}{##1}}}
\expandafter\def\csname PYGdefault@tok@dl\endcsname{\def\PYGdefault@tc##1{\textcolor[rgb]{0.73,0.13,0.13}{##1}}}
\expandafter\def\csname PYGdefault@tok@s2\endcsname{\def\PYGdefault@tc##1{\textcolor[rgb]{0.73,0.13,0.13}{##1}}}
\expandafter\def\csname PYGdefault@tok@sh\endcsname{\def\PYGdefault@tc##1{\textcolor[rgb]{0.73,0.13,0.13}{##1}}}
\expandafter\def\csname PYGdefault@tok@s1\endcsname{\def\PYGdefault@tc##1{\textcolor[rgb]{0.73,0.13,0.13}{##1}}}
\expandafter\def\csname PYGdefault@tok@mb\endcsname{\def\PYGdefault@tc##1{\textcolor[rgb]{0.40,0.40,0.40}{##1}}}
\expandafter\def\csname PYGdefault@tok@mf\endcsname{\def\PYGdefault@tc##1{\textcolor[rgb]{0.40,0.40,0.40}{##1}}}
\expandafter\def\csname PYGdefault@tok@mh\endcsname{\def\PYGdefault@tc##1{\textcolor[rgb]{0.40,0.40,0.40}{##1}}}
\expandafter\def\csname PYGdefault@tok@mi\endcsname{\def\PYGdefault@tc##1{\textcolor[rgb]{0.40,0.40,0.40}{##1}}}
\expandafter\def\csname PYGdefault@tok@il\endcsname{\def\PYGdefault@tc##1{\textcolor[rgb]{0.40,0.40,0.40}{##1}}}
\expandafter\def\csname PYGdefault@tok@mo\endcsname{\def\PYGdefault@tc##1{\textcolor[rgb]{0.40,0.40,0.40}{##1}}}
\expandafter\def\csname PYGdefault@tok@ch\endcsname{\let\PYGdefault@it=\textit\def\PYGdefault@tc##1{\textcolor[rgb]{0.25,0.50,0.50}{##1}}}
\expandafter\def\csname PYGdefault@tok@cm\endcsname{\let\PYGdefault@it=\textit\def\PYGdefault@tc##1{\textcolor[rgb]{0.25,0.50,0.50}{##1}}}
\expandafter\def\csname PYGdefault@tok@cpf\endcsname{\let\PYGdefault@it=\textit\def\PYGdefault@tc##1{\textcolor[rgb]{0.25,0.50,0.50}{##1}}}
\expandafter\def\csname PYGdefault@tok@c1\endcsname{\let\PYGdefault@it=\textit\def\PYGdefault@tc##1{\textcolor[rgb]{0.25,0.50,0.50}{##1}}}
\expandafter\def\csname PYGdefault@tok@cs\endcsname{\let\PYGdefault@it=\textit\def\PYGdefault@tc##1{\textcolor[rgb]{0.25,0.50,0.50}{##1}}}

\makeatother

\begin{document}
\title{\tool: A Toolkit to Naturalize the Source Code Corpus}


\author{\IEEEauthorblockN{Yao Wan$^1$, Yang He$^2$, Jian-Guo Zhang$^3$, \\Yulei Sui$^2$, Hai Jin$^1$, Guandong Xu$^2$, Caiming Xiong$^4$, Philip S. Yu$^3$}
	\IEEEauthorblockA{$^1$Services Computing Technology and System Lab, Cluster and Grid Computing Lab, \\School of Computer Science and Technology, Huazhong University of Science and Technology, Wuhan, China\\
	$^2$School of Computer Science, University of Technology Sydney, NSW, Australia \\
	$^3$Department of Computer Science, University of Illinois at Chicago, Illinois, USA\\
	$^4$Salesforce Research, Palo Alto, USA\\
	\{wanyao,hjin\}@hust.edu.cn, \{yulei.sui,guandong.xu\}@uts.edu.au, \{jzhan51,psyu\}@uic.edu, cxiong@salesforce.com}
}
\maketitle

\begin{abstract}
	We present \tool, an efficient and extensible toolkit to bridge the gap between natural language and programming language, and facilitate the research on big code analysis.
	Using \tool, researchers both from natural language or programming language communities can quickly and easily reproduce the state-of-the-art baselines and implement their  approach.
	\tool is built upon Fairseq and PyTorch, providing (1) an efficient computation with multi-GPU and mixed-precision data processing for fast model training, (2) a modular and extensible framework that makes it easy to reproduce or implement an approach for big code analysis, and (3) a command line interface and a graphical user interface to demonstrate each model's performance. 
	Currently, we have included several state-of-the-art baselines across different tasks (e.g., code completion, code comment generation, and code retrieval) for demonstration. 
	The video of this demo is available at \texttt{\url{https://www.youtube.com/watch?v=q4W5VSI-u3E&t=25s}}.

\end{abstract}


\section{Introduction}
The rapid growth of machine learning (ML), especially of deep learning (DL) based natural language processing (NLP), brings great opportunities to explore and exploit NLP techniques for various tasks of software engineering (SE), e.g., code documentation~\cite{wan2018improving,alon2018code2seq}, code completion~\cite{kim2020code} and code retrieval~\cite{gu2018deep,wan2019multi-modal}. 
The underlying insights for learning-based code analysis is the naturalness hypothesis shared among natural languages and programming languages.
Despite the flourishing study, many state-of-the-art approaches have been  suffering from the replicability and reproducibility issues. This is due to the fact that the performance of deep learning approaches is sensitive to  datasets and hyperparameters, and currently, no unified open-source toolkit is available to our communities.

To fill this gap, this paper introduces \tool,\footnote{The term NaturalCC also represents natural code comprehension, which is a fundamental task lies in the synergy between the programming language and NLP.} a comprehensive platform to facilitate research NLP-based big code analysis.
Both of the researchers from SE community or NLP community can benefit from the toolkit for fast training and reproduction. \tool features the following advantages:
\begin{itemize}
	\item \textbf{Efficient Data Preprocessing.} We have cleaned and preprocessed three public datasets (i.e., CodeSearchNet~\cite{husain2019codesearchnet}, Py150~\cite{py150}, and Python~\cite{wan2018improving}) for different tasks. All the data loaders and processes in our model training can be parallelized in multiple GPUs. Besides, our toolkit also supports the mixed-precision for numerical calculation.
	\item \textbf{Extensibility and Modularity.} Based on the registry mechanism implemented in Fairseq~\cite{ott2019fairseq}, our framework is well modularized and can be easily extended to various tasks. In particular, when implementing a new task, we only need to implement the task and models in the corresponding folders and then register them. 
	\item \textbf{Flexible Interface.} We provide flexible APIs for developers to easily invoke the trained models for other applications.
\end{itemize}

Additionally, we demonstrate \tool with a command line interface as well as a graphical user interface, using three application tasks, i.e., code completion, code comment generation, and code retrieval.

\tool is an ongoing open-source toolkit maintained by the \textit{CodeMind} team. We hope \tool can facilitate the research of software engineering with natural language processing. We also encourage researchers to integrate their state-of-the-art approach into \tool, to promote the research in both communities.

All the source code and materials are publicly available via GitHub: \texttt{\url{http://github.com/CGCL-codes/naturalcc}}.\footnote{The open-source Fairseq toolkit has inspired us a lot, and our open-source project also follows the MIT license.}
We also build a website for our team and will post the updates in \texttt{\url{http://xcodemind.github.io}}.

\section{Architecture Design}
\begin{figure}[t!]
	\centering
	\includegraphics[width=0.485\textwidth]{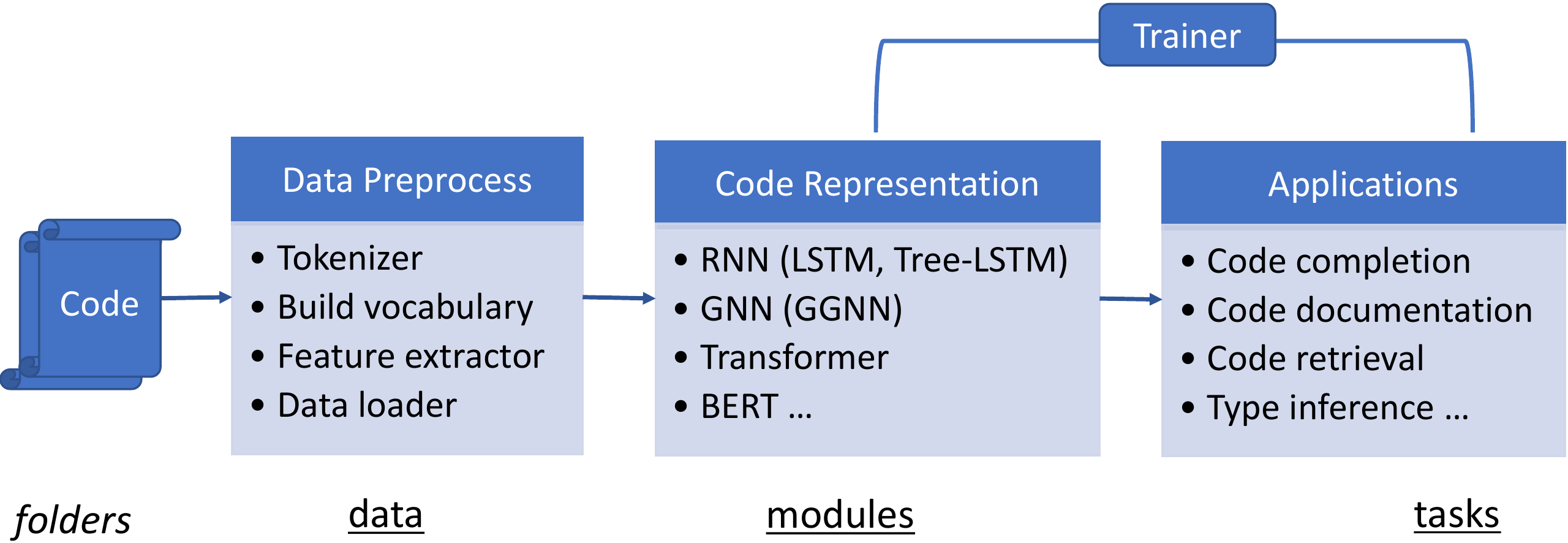}
	\caption{A pipeline of \tool}
	\label{fig_architecture}
\end{figure}

Figure~\ref{fig_architecture} shows a pipeline of our \tool. Given a dataset of code snippets, we first preprocess the data in the data preprocessing stage and then feed each mini-batch of samples into the code representation module, a fundamental component for several downstream tasks. In the code representation module, we have implemented many state-of-the-art encoders (e.g., RNN~\cite{hochreiter1997long}, GNN~\cite{wu2020comprehensive}, Transformer~\cite{vaswani2017attention} and BERT~\cite{devlin2018bert}). Based on the code representation, \tool can also support various downstream tasks, e.g., code documentation, programming language modeling, code retrieval and type inference. The designed \texttt{Trainer} controls model training for each task.

\subsection{Data Preprocessing}
In the data preprocessing stage, we first tokenize the source code by a tokenizer (e.g., space tokenizer or BPE~\cite{karampatsis2020big} tokenizer) and then build a vocabulary for these tokens. In addition, we can also extract some domain-specific features (e.g., AST~\cite{wan2018improving}, control-flow graphs~\cite{wan2019multi-modal}, or data-flow graphs~\cite{allamanis2017learning}). The goal of this process is to build a series of mini-batches for training. We put all the data-related processes in the \texttt{data} and \texttt{dataset} folders. 
\subsection{Code Representation}
Code representation/understanding, which aims to learn an embedding vector, is one of the most critical components for big code analysis. In \tool, we have included most state-of-the-art neural network encoders to represent the source code and their extracted features. For example, we have implemented RNN-based models to represent the sequential tokens or (linearized) AST of code. We implement graph neural networks (GNNs) such as gated graph neural networks (GGNNs) to represent the graph structure features of code (e.g., control-flow and data-flow graphs). We have also included the advanced Transformer networks~\cite{vaswani2017attention}, which serve as the replacement of the RNN network, with its fast computation and ability to handle long-range dependent sequence. In addition, the \tool also supports the masked pre-trained models, e.g., BERT and RoBERTa~\cite{liu2019roberta}. We put all the code representation networks in the \texttt{models} and \texttt{modules} folders. 

\subsection{Applications}
Our \tool supports many different downstream tasks. We have currently implemented three tasks, i.e., code completion, code comment generation, and code retrieval, to validate the effectiveness of the proposed framework. All the referred baselines on each task have been carefully checked and evaluated when compared against the released source code from the original papers. The implementations and the tasks in this toolkit will serve as baselines for fair comparison for future research use. We organize all the tasks in the \texttt{tasks} folder.

\subsubsection{Code Completion}
Code completion, which predicts the next code element based on the previously written code, has become an essential tool in many IDEs. It can boost developers' programming productivity. In this task, we have implemented the referred model SeqRNN~\cite{raychev2014code} and TravTrans~\cite{kim2020code}.
We train the models in the Py150 dataset and evaluate them using the MRR metrics.

\subsubsection{Code Comment Generation}
Generating comments for code snippets is an effective way for program  understanding and facilitate the software development and maintenance. In this task, we have implemented the referred model NeuralTransformer~\cite{ahmad2020transformer}.
We trained the model in Python and Java datasets and evaluated them using the BLEU and Rouge metrics. 

\subsubsection{Code Retrieval} 
Searching semantically similar code snippets given a natural language query can provide developers a series of templates as reference for rapid  prototyping.
In this task, we have implemented four benchmark baselines (i.e., NBOW, 1D-CNN, biRNN and SelfAttn) and evaluated them by using the MRR metrics on CodeSearchNet dataset~\cite{husain2019codesearchnet}.

Additional tasks with state-of-the-art models are still under development. They will be released soon, including code clone detection~\cite{hua2019fcca}, type inference~\cite{hellendoorn2018deep}, vulnerability detection~\cite{li2018vuldeepecker}, and masked language modeling for code pre-training~\cite{feng2020codebert}.
\subsection{Trainer}
We have designed a trainer (\texttt{ncc\_trainer.py}) module to control the whole training process of models.  Furthermore, we have designed and implemented a simpler trainer in a universal way  (\texttt{ncc\_trainer\_simple.py}) for those beginners who are not familiar with our framework.

\section{Implementation}
We have implemented \tool based on the Fairseq and PyTorch. Following the outstanding registry mechanism designed in Fairseq, \tool has good extensibility with the modularized design. 

\subsection{Registry Mechanism}

\begin{listing}[H]
\caption{The registry mechanism in \texttt{\_\_init\_\_.py}} 
\begin{Verbatim}[commandchars=\\\{\},xleftmargin=2em,fontsize=\footnotesize,breaklines=true,linenos]
\PYG{k}{def} \PYG{n+nf}{register\PYGZus{}task}\PYG{p}{(}\PYG{n}{name}\PYG{p}{):}
  \PYG{k}{def} \PYG{n+nf}{register\PYGZus{}task\PYGZus{}cls}\PYG{p}{(}\PYG{n+nb+bp}{cls}\PYG{p}{):}
    \PYG{k}{if} \PYG{n}{name} \PYG{o+ow}{in} \PYG{n}{TASK\PYGZus{}REGISTRY}\PYG{p}{:}
      \PYG{k}{raise} \PYG{n+ne}{ValueError}\PYG{p}{(}\PYG{l+s+s1}{\PYGZsq{}Duplicate error...\PYGZsq{}}\PYG{p}{)}
    \PYG{n}{TASK\PYGZus{}REGISTRY}\PYG{p}{[}\PYG{n}{name}\PYG{p}{]} \PYG{o}{=} \PYG{n+nb+bp}{cls}
    \PYG{n}{TASK\PYGZus{}CLASS\PYGZus{}NAMES}\PYG{o}{.}\PYG{n}{add}\PYG{p}{(}\PYG{n+nb+bp}{cls}\PYG{o}{.}\PYG{n+nv+vm}{\PYGZus{}\PYGZus{}name\PYGZus{}\PYGZus{}}\PYG{p}{)}
    \PYG{k}{return} \PYG{n+nb+bp}{cls}
    
  \PYG{k}{return} \PYG{n}{register\PYGZus{}task\PYGZus{}cls}
\end{Verbatim}
\end{listing}
We have implemented a \texttt{register} decorator in the entry of building a task, model or module (cf. \texttt{\_\_init\_\_.py} in each folder). Listing~1 shows the workflow of registering a new task. In brief, the registry mechanism is to design a global variable to store each task of model objects for off-the-shelf fetching.
This registry mechanism can provide us the ability of extension, as we only need to include this decorator when defining a new task/model/module in the corresponding function. 

\subsection{Multi-GPU Training}
Following Fairseq, we use the \texttt{NCCL} library and \texttt{torch.distributed} to support model training  on multiple GPUs. Every GPU stores a copy of model parameters, and the global optimizer functions as  synchronous optimization in each GPU. Gradients accumulation is also supported to mitigate multi-GPU computation lagging.

\subsection{Mixed-Precision}
\tool can also support both full precision (FP32) and half-precision floating point (FP16) for fast training and inference. From our experience, setting the FP16 option can largely reduce the memory usage, and further save the training time. To preserve model accuracy, the  parameters are stored in FP32 while updated by FP16 gradients.

\subsection{An Implementation Example}
We take code completion as an example to show the pipeline of how to quickly build a new task in \tool. Note that, in this section, we only describe the main steps in each file, and more details are referred to the corresponding source files.

\textbf{\textit{Building a task.}} In the first step, we create a \texttt{CompletionTask} in the \texttt{ncc/tasks/completion.py}, with a decorator \texttt{register\_task} around. Listing~2 shows the whole processing of building a new task.
This \texttt{class} provides a function \texttt{build\_model} for building a model according to the arguments defined by users.


\begin{listing}[H]
\caption{\texttt{tasks/completion/completion.py}}
\begin{Verbatim}[commandchars=\\\{\},xleftmargin=2em,fontsize=\footnotesize,breaklines=true,linenos]
\PYG{n+nd}{@register\PYGZus{}task}\PYG{p}{(}\PYG{l+s+s1}{\PYGZsq{}completion\PYGZsq{}}\PYG{p}{)}
\PYG{k}{class} \PYG{n+nc}{CompletionTask}\PYG{p}{(}\PYG{n}{NccTask}\PYG{p}{):}
  \PYG{k}{def} \PYG{n+nf+fm}{\PYGZus{}\PYGZus{}init\PYGZus{}\PYGZus{}}\PYG{p}{(}\PYG{n+nb+bp}{self}\PYG{p}{,} \PYG{n}{args}\PYG{p}{,} \PYG{n}{dictionary}\PYG{p}{):}
    \PYG{n+nb}{super}\PYG{p}{()}\PYG{o}{.}\PYG{n+nf+fm}{\PYGZus{}\PYGZus{}init\PYGZus{}\PYGZus{}}\PYG{p}{(}\PYG{n}{args}\PYG{p}{)}
    \PYG{n+nb+bp}{self}\PYG{o}{.}\PYG{n}{dictionary} \PYG{o}{=} \PYG{n}{dictionary}

  \PYG{n+nd}{@classmethod}
  \PYG{k}{def} \PYG{n+nf}{setup\PYGZus{}task}\PYG{p}{(}\PYG{n+nb+bp}{cls}\PYG{p}{,} \PYG{n}{args}\PYG{p}{,} \PYG{o}{**}\PYG{n}{kwargs}\PYG{p}{):}
    \PYG{n}{dictionary} \PYG{o}{=} \PYG{n+nb+bp}{cls}\PYG{o}{.}\PYG{n}{load\PYGZus{}dictionary}\PYG{p}{(}\PYG{n}{args}\PYG{p}{)}
    \PYG{k}{return} \PYG{n+nb+bp}{cls}\PYG{p}{(}\PYG{n}{args}\PYG{p}{,} \PYG{n}{dictionary}\PYG{p}{)}

  \PYG{k}{def} \PYG{n+nf}{build\PYGZus{}model}\PYG{p}{(}\PYG{n+nb+bp}{self}\PYG{p}{,} \PYG{n}{args}\PYG{p}{):}
    \PYG{n}{model} \PYG{o}{=} \PYG{n+nb}{super}\PYG{p}{()}\PYG{o}{.}\PYG{n}{build\PYGZus{}model}\PYG{p}{(}\PYG{n}{args}\PYG{p}{)}
    \PYG{k}{return} \PYG{n}{model}
\end{Verbatim}
\end{listing}

\textbf{\textit{Building a model.}} Listing~3 shows the process of building an  RNN model for code completion. We define a new class \texttt{SeqRNNModel} in the \texttt{ncc/models/completion/seqrnn.py}, which inherits the \texttt{NccLanguageModel}. In this class, we build a decoder network \texttt{LSTMDecoder}, which is implemented in the \texttt{modules} folder.


\begin{listing}[H]
\caption{\texttt{models/completion/seqrnn.py}}
\begin{Verbatim}[commandchars=\\\{\},xleftmargin=2em,fontsize=\footnotesize,breaklines=true,linenos]
\PYG{n+nd}{@register\PYGZus{}model}\PYG{p}{(}\PYG{l+s+s1}{\PYGZsq{}seqrnn\PYGZsq{}}\PYG{p}{)}
\PYG{k}{class} \PYG{n+nc}{SeqRNNModel}\PYG{p}{(}\PYG{n}{NccLanguageModel}\PYG{p}{):}
  \PYG{k}{def} \PYG{n+nf+fm}{\PYGZus{}\PYGZus{}init\PYGZus{}\PYGZus{}}\PYG{p}{(}\PYG{n+nb+bp}{self}\PYG{p}{,} \PYG{n}{args}\PYG{p}{,} \PYG{n}{decoder}\PYG{p}{):}
    \PYG{n+nb}{super}\PYG{p}{()}\PYG{o}{.}\PYG{n+nf+fm}{\PYGZus{}\PYGZus{}init\PYGZus{}\PYGZus{}}\PYG{p}{(}\PYG{n}{decoder}\PYG{p}{)}
    \PYG{n+nb+bp}{self}\PYG{o}{.}\PYG{n}{args} \PYG{o}{=} \PYG{n}{args}

  \PYG{n+nd}{@classmethod}
  \PYG{k}{def} \PYG{n+nf}{build\PYGZus{}model}\PYG{p}{(}\PYG{n+nb+bp}{cls}\PYG{p}{,} \PYG{n}{args}\PYG{p}{,} \PYG{n}{config}\PYG{p}{,} \PYG{n}{task}\PYG{p}{):}
    \PYG{n}{decoder} \PYG{o}{=} \PYG{n}{LSTMDecoder}\PYG{p}{(}
      \PYG{o}{...}
    \PYG{p}{)}
    \PYG{k}{return} \PYG{n+nb+bp}{cls}\PYG{p}{(}\PYG{n}{args}\PYG{p}{,} \PYG{n}{decoder}\PYG{p}{)}
\end{Verbatim}
\end{listing}

\textbf{\textit{Model training.}} Listing~4 shows the construction of a \texttt{Trainer} and the pipeline of train steps. Core parameters are involved in this \texttt{class} such that pre-trained models can be precisely restored during inference or fine-tuning. 


\begin{listing}[H]
\caption{\texttt{trainer/trainer.py}}
\begin{Verbatim}[commandchars=\\\{\},xleftmargin=2em,fontsize=\footnotesize,breaklines=true,linenos]
\PYG{c+c1}{\PYGZsh{} 1. Setup task, e.g., completion, comment generation, etc.}
\PYG{n}{task} \PYG{o}{=} \PYG{n}{tasks}\PYG{o}{.}\PYG{n}{setup\PYGZus{}task}\PYG{p}{(}\PYG{n}{args}\PYG{p}{)}
\PYG{c+c1}{\PYGZsh{} 2. Build model and criterion}
\PYG{n}{model} \PYG{o}{=} \PYG{n}{task}\PYG{o}{.}\PYG{n}{build\PYGZus{}model}\PYG{p}{(}\PYG{n}{args}\PYG{p}{)}
\PYG{n}{criterion} \PYG{o}{=} \PYG{n}{task}\PYG{o}{.}\PYG{n}{build\PYGZus{}criterion}\PYG{p}{(}\PYG{n}{args}\PYG{p}{)}
\PYG{c+c1}{\PYGZsh{} 3. Build trainer}
\PYG{n}{trainer} \PYG{o}{=} \PYG{n}{Trainer}\PYG{p}{(}\PYG{n}{args}\PYG{p}{,} \PYG{n}{task}\PYG{p}{,} \PYG{n}{model}\PYG{p}{,} \PYG{n}{criterion}\PYG{p}{)}
\PYG{k}{while} \PYG{p}{(}
  \PYG{n}{lr} \PYG{o}{\PYGZgt{}} \PYG{n}{args}\PYG{p}{[}\PYG{l+s+s1}{\PYGZsq{}optimization\PYGZsq{}}\PYG{p}{][}\PYG{l+s+s1}{\PYGZsq{}min\PYGZus{}lr\PYGZsq{}}\PYG{p}{]}
  \PYG{o+ow}{and} \PYG{n}{epoch\PYGZus{}itr}\PYG{o}{.}\PYG{n}{next\PYGZus{}epoch\PYGZus{}idx} \PYG{o}{\PYGZlt{}=} \PYG{n}{max\PYGZus{}epoch}
  \PYG{o+ow}{and} \PYG{n}{trainer}\PYG{o}{.}\PYG{n}{get\PYGZus{}num\PYGZus{}updates}\PYG{p}{()} \PYG{o}{\PYGZlt{}} \PYG{n}{max\PYGZus{}update}
\PYG{p}{):}
  \PYG{n}{task}\PYG{o}{.}\PYG{n}{train\PYGZus{}step}\PYG{p}{(}\PYG{n}{samples}\PYG{p}{)}
\end{Verbatim}
\end{listing}

\section{Demonstration}
\subsection{Command Line Interface}
\tool provides a command line interface that enables researchers and developers to simply explore the included state-of-the-art models. For each code analysis related tasks, users can try this command:

\begin{Verbatim}[commandchars=\\\{\},xleftmargin=2em,fontsize=\footnotesize,breaklines=true,linenos]
\PYG{n+nv}{\PYGZdl{}python} \PYGZhy{}m cli.predictor \PYGZhy{}m \PYGZlt{}model\PYGZgt{} \PYGZhy{}i \PYGZlt{}input\PYGZgt{}
\end{Verbatim}
where \texttt{-m} is the pre-trained model directory, and \texttt{-i} is the  corresponding user input (a partial code snippet in the code completion task). \tool will automatically load model parameters,  process the user input and return inference information in details.

\begin{figure*}[t!]
	\centering
	\includegraphics[width=0.91\textwidth]{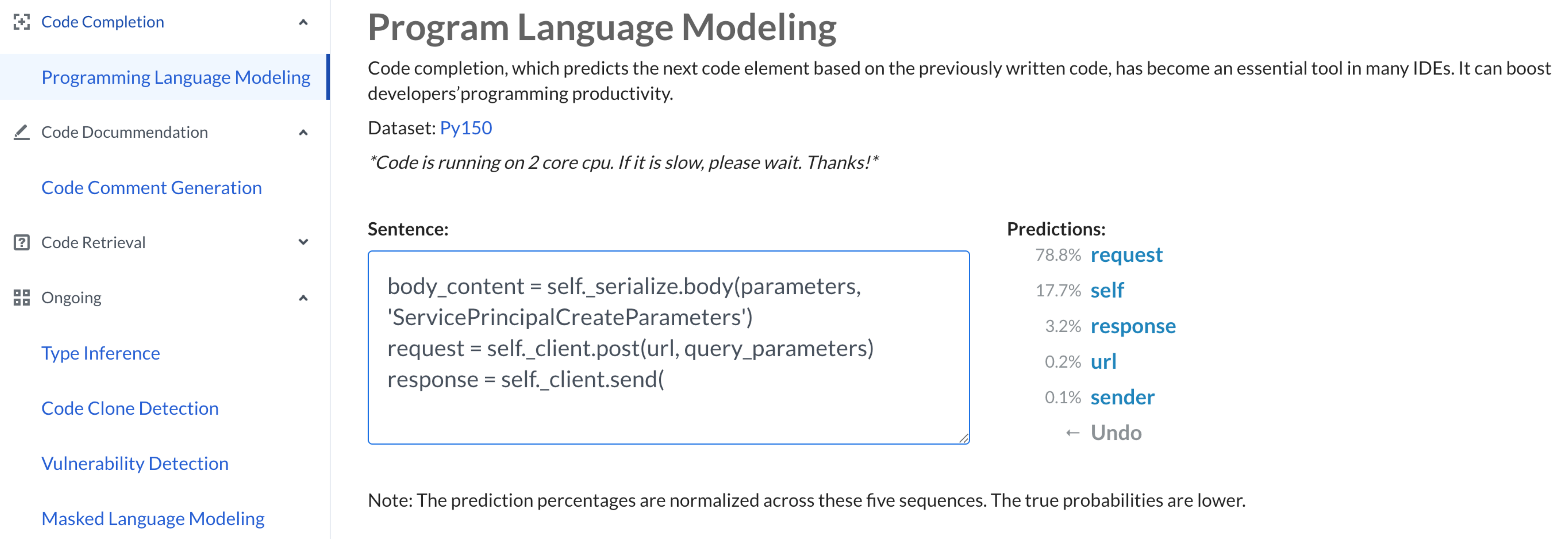}
	\caption{A screenshot of our graphical user interface for demonstration}
	\label{top_demo_gui}
\end{figure*}

\subsection{Graphical User Interface}
We have also provided a graphical user interface for users to easily access and explore each trained model's results through an online Web browser.
The design of our Web is based on the open source demo of AllenNLP~\cite{Gardner2017AllenNLP}.
We have deployed the graphical demo in the Nginx server and provided flexible APIs via the Flask engine.

As shown in Figure~\ref{top_demo_gui}, we have integrated three popular software engineering tasks for demonstration, i.e., code completion, code comment generation, and code retrieval.
Taking code completion as an example. By default, we have implemented this task based on the programming language modeling. 
Given a series of written tokens by Python, the predicted tokens with corresponding probabilities generated by our model will appear simultaneously when the user enters the next tokens.
In this page, the users can also select the trained model accordingly.
\section{Conclusion and Future Work}
This paper presents \tool, an efficient and extensible open-source toolkit to bridge the gap between the programming language and natural language. Currently, \tool has implemented several state-of-the-art models across three popular software engineering tasks. We provide a detailed sample as an example to quickly implement a new task. For demonstration, we have provided a command line tool as well as a graphical user interface for other researchers to do quick prototyping.
All the materials about this toolkit can be accessed from \texttt{\url{http://xcodemind.github.io}}.

We will extend this toolkit to more software engineering tasks in our future work, including code clone detection, vulnerability detection, and masked language modeling. We also encourage more researchers to join our team to promote the development of this toolkit as well as the whole research community.

\section*{Acknowledgements}
The Fairseq highly inspires the \tool. We appreciate the Fairseq team for their contribution and the high-quality backbone structure of the framework.
\bibliographystyle{IEEEtran}
\bibliography{ref}

\end{document}